\documentstyle{article}
\begin{document}
\title{Landau Theory of  the Mott Transition   in the Fully Frustrated Hubbard
Model in   Infinite Dimensions}
\author{ Gabriel Kotliar \\
Serin Physics Laboratory,   Rutgers University, \\
Piscataway, NJ 08854, USA}
\date{\today}
\maketitle
\begin{abstract}
We discuss   the  solution of
the Mott transition
problem in a fully frustrated lattice with a semicircular density of
states in the limit of infinite dimensions from
the point of view of a Landau free energy functional.
This approach provides a simple relation
between  the free   energy  of
the lattice model and that of its local
description in terms of an  impurity model.
The character of the Mott transition in infinite dimensions,
(as reviewed by Georges Kotliar Krauth and Rozenberg, RMP 68, 1996, 13)
follows simply  from the form of the free
energy  functional  and the
physics of quantum impurity models.
At zero temperature, below a critical value of the interaction U,
a     Mott insulator with a finite gap in the
one particle spectrum,  becomes  unstable to the formation
of a narrow band near the Fermi energy.
Using the insights provided by the Landau approach  we  answer 
questions  raised about the dynamical
mean field solution of the Mott transition problem, 
and comment on its 
applicability
to  three dimensional transition metal oxides.
\end{abstract}
\section{Introduction}
\def\den{({\varepsilon_{k'}} - {\varepsilon_{k}})}
\def\tree{ab}
\def\num{{{V_{k}}^2
{V_{k'}}^2}}
\def\num1{{{\Vkl}^2
{\Vklp}^2}}

The idea of understanding lattice models of correlated electrons
from a local perspective is a very intuitive one,  and has been used
repeatedly in many body physics  over the years.

\newcommand{\da}{\downarrow}
\newcommand{\ua}{\uparrow}

A well known   example is the heavy fermion problem, where a broad band
of conduction electrons interacts  with more localized f electrons,
via a magnetic Kondo exchange interaction.
In the early  days of the heavy fermion problem,  a great deal
of understanding was obtained by considering the screening of an
isolated
spin  by a sea of conduction electrons, i.e. studying  the single site Kondo effect,
and then regarding the Kondo lattice as a collection of Kondo
impurity models.

This view had some successes in   explaining   the origin of
a non perturbative energy scale, the Kondo temperature,
where
the properties of the system change dramatically (for instance the
susceptibility crosses over from Curie to Pauli  behavior ).

In an early  paper, however,
 Nozi\`eres \cite{nozieres1} pointed out, that 
the physics of  
the lattice problem
is   more complex than the single impurity problem
since at least in the limit of low density of conduction electrons,
there are not enough itinerant    electron spins,   to screen all the  impurity
spins in the lattice. In this case
one cannot regard the lattice as a collection of single Kondo impurities.
This issue is now known as Nozi\`eres'  exhaustion problem.
This is perhaps one of the earliest
warnings that   single impurity  thinking can be misleading if it
is applied uncritically  to lattice  problems involving a correlated
degree of freedom at each site.

In the context of transition metal oxide physics, an impurity view
of the  d-electron spectral function was put forward by Zaanen,
Sawatzky and Allen \cite{allen} 
and by Fujimori, Minami and Sugano\cite{fujimori}, and led
to a qualitative description of the spectra in these systems.
In the light of the modern developments of the  dynamical mean field
theory,  we would regard the early applications of impurity  views
to the physics of f and d electron systems
as local  (but not self consistent)
impurity approximations to lattice models.

The last ten years have witnessed dramatic progress in the theory
of correlated electron systems.
The modern developments of  a dynamical mean field theory \cite{review}
and its
implementation via mappings onto impurity models
\cite{ipt,mj1}, now allows  us
to  use impurity models supplemented by a self consistency
condition to study lattice models.  The results are exact in
the well defined limit of infinite lattice coordination \cite{vollhardt}.

We are now in a much better position  to  gauge
the reliability of the arguments based on the
Local Impurity Approximation by studying lattice models in the limit
of large lattice coordination using
Local Impurity  {\it Self-Consistent}
Approximations. If the self consistency condition does not play
an important role, naive impurity  based arguments are reliable.
Since the dynamical mean field theory is exact in the limit of large
lattice coordination, we can also understand which physical elements
are absent in this limit (a most notable example is the feedback of  
the magnetic
correlations on the single particle properties) 
and  assess in which
physical circumstances it provides reliable guidance to the physics
of three dimensional real  materials.

This paper is devoted to the problem of the Mott transition, i.e. the
interaction driven 
metal insulator transition,  and its description in terms
of quantum impurity problems.
We consider   
a  half filled Hubbard model on a
fully frustrated lattice
\cite{g1}
with  a semicircular density of states
in the limit of infinite lattice coordination.
The term frustration 
refers to the degree of magnetic frustration in the  parent Mott insulator.

Ref  \cite{g1}  reported that   this model
exhibits  a Mott transition at a critical  value of the ratio $ U \over t$.
The correct  description of  the destruction of the metallic state
at zero temperature as the  Mott
transition  is approached by increasing  
U, was proposed by Zhang, Rozenberg and Kotliar (ZRK) \cite{zrk}
on the basis
of the iterated perturbation theory  (IPT) \cite{ipt}.

The calculations of ZRK  revealed  that 
while the approach to the Mott
transition  from the metallic side is driven by a collapse
of the Fermi energy, as in Brinkman Rice \cite{rice} theory,
it also exhibits new unusual features.
The metal disappears into an insulator with
a preformed (finite) Mott Hubbard gap. We will refer to
this  as the semicontinuous 
scenario to be distinguished from the  competing, bicontinous, scenario where 
the gap closes at the same point where the quasiparticle
weight vanishes  as   discussed on  page \pageref{bicon}.

The complete picture, of the Mott Hubbard transition in infinite
dimensions emerged with the  work  of A. Georges and W. Krauth   \cite{werner}
\cite{werner2}
and  Rozenberg et. al.  \cite{g2} \cite{g3},
who described the
destruction of the insulating state at zero temperature, the first order finite 
temperature metal insulator transition, and the  crossovers
that govern the behavior above the finite temperature second order  
critical  point. 
They produced  a wealth of  physical results, which
were in surprisingly good  qualitative  agreement with experimental
data.
\cite{phase}.
The zero temperature scenario for the destruction
of the metallic state was  put on a   firm  footing
by  the development
of the projective self consistent method \cite{moeller1}. This method
overcame   the difficulties associated with the presence of
several energy scales,  which had beset 
earlier  treatments.

In spite of these developments, several  
questions   about the solution of the Hubbard model
in large dimensions were   raised
\cite{gep}  
\cite{logan} 
\cite{nozieres}
\cite{kehrein}
and  numerical
studies were undertaken  in an attempt to answer them. 
\cite{dresden}.  
\cite{schlipf}
\cite{bulla}
This renewed interest and in particular the insightful questions of Nozi\`eres \cite{nozieres},
motivates  us to reexamine
the problem from a new perspective, that of a Landau -like free energy
functional of a  "metallic order parameter", generalizing an 
approach used in our earlier studies of  interacting random models  with V.
Dobrosavlevic \cite{vlad}.

Our discussion highlights the peculiar character
of the Landau theory of the metal to insulator transition.
This singular dependence of the  mean field free energy
on the metallic order parameter (and not the specific approximations
such as IPT, QMC or exact diagonalization of finite systems which
were used in the early studies of this problem)
is responsible for
the unusual features of
the solution of the Hubbard model in infinite dimensions.

Our aim is partly   pedagogical,
we  use  the Landau functional 
to describe 
from a  new perspective
results that were obtained 
a  few years ago.
Besides clarifying
the existing   confusion in the literature 
of
the subject,
there is another purpose in
writing a pedagogical note. There are not that many  solvable
models of the Mott transition in dimensions higher than one!
We believe that there are still many lessons to be drawn from the solution
in the limit of infinite dimensions, 
that can be of  use  in  tackling  more difficult problems, 
in the field  of  strongly  correlated electron systems. 
We believe that
the Landau-like approach  
which we advocate in this paper 
can be valuable in other dynamical mean field studies.
Finally, while we believe that  the nature 
of the Mott transition in fully frustrated
systems  in the limit of infinite dimensions,
has been   understood at the qualitative level,
there  still remains a large amount of
quantitative
work to be done   on this  problem. Our insights
should be a helpful guide to further  investigations.
\def\Vkl{V^{L}_{k}}
\def\Vklp{V^{L}_{k^\prime}}

This paper is organized as follows.
After setting the notation in section \ref{notation},
summarizing the  scenario describing the destruction of the  metallic state  
in section \ref{zrk}. 
We state the questions raised  
by this suggestion
and  describe the alternative (bicontinous) scenario where the gap closes 
at the same point where the quasiparticle peak disappears  in section \ref{criticism}.
Two technical tools are essential  to justify   the
validity of the semicontinuous scenario, the Landau
free energy functional is described in section \ref{functional},
and some results of the  projective self consistent method 
are summarized in section \ref{projectivesec}.
Using  these tools, we  describe the energetics of the
metal insulator transition, inspired by the questions of 
Nozi\`eres
\cite{logan}\cite{nozieres}. The Landau functional, provides us with
a concrete bridge between the impurity model and the lattice
model allowing us to use our  knowledge of Kondo
impurity  physics  to 
understand the Mott transition problem.

In section \ref{uc1sec}.
we  use the Landau functional to describe 
 the arguments of Fisher,  Kotliar and Moeller 
\cite{moeller2}
for the   determination of   the conditions for $ U_{c1}$,
the point where the insulator disappears.
Near $ U_{c1}$ 
the  physical picture is that
of an  impurity  in a weakly coupled regime,
Nozi\`eres exhaustion ideas are applicable in this case. 

In section 
\ref{uc2sec} we recall
the arguments of Moeller et. al.
\cite{moeller1}  for the disappearance
of the metal at  the critical value  $U_{c2}$.
Here,
the Mott insulator with a finite gap is indeed
unstable towards the formation of a  narrow metallic 
band at the Fermi level.
The effective impurity description
is in an  intermediate coupling regime.
From the perspective of  our analysis based on a   Landau  functional,
the semicontinous scenario, i.e. the fact that $U_{c1} < U_{c2}$, is
an unavoidable consequence of the  different behaviors of quantum
impurity models in   weak   and strong coupling limits. 

In section \ref{conclusions}
we argue that  a more realistic  consideration of the  
magnetic correlations
in finite dimension,
may change the character of
the free energy functional 
and comment on the relevance of the dynamical mean field
theory  results to 
finite dimensional systems.

\section{Lattice Model and Associated Impurity Hamiltonian}
\label{notation}
We consider the Hubbard model on the  Bethe lattice
in the paramagnetic phase  with
coordination d and hopping $\frac{t}{\sqrt{d}}$   at half filling.
$$H = - \sum_{ <i,j> \sigma } {\frac{t}{\sqrt{d}}} (c^{+}_{i \sigma}
c_{j \sigma} + c.c.) + \sum_{i} U n_{i \uparrow }n_{i \downarrow}
$$
The half bandwidth is given by  $D = 2 t$ 
and we will use
$D=1$ as a unit of energy.
The kinetic energy  per site, K, can always be expressed in terms of the non local
Green's function $G_{i,j} $. In the limit of large lattice coordination
it can also be expressed in terms of the one particle
Green's function:
\def\ee{e^{i \omega 0^+}}
\begin{equation}
 <K > =    -{\frac{1}{N_{s}} } \sum_{\sigma, <i,j> \omega}  \ee G_{i,j}(i \omega)  t_{i,j}  = t^{2}
 2   T \sum_{\omega} G (i \omega)^{2}
\label{kinetic1}
\end{equation}
In the limit of large dimensions the total energy per site,  
$E={<H> \over N_{s}}$ 
reduces to:

\begin{equation}
E =  T \sum_{\omega}{ [(i \omega + \mu ) G(i \omega) -1 ] 
e^{i \omega 0^+}} + {\frac{1}{2}} < K >
\label{total}
\end{equation}
The interaction energy per site is given by:
\begin{equation}
< U >  = E  - < K > = U < n_{i \uparrow }n_{i \downarrow}>
\end{equation}
As is well known now \cite{ipt},
all the local correlation functions of the model
can be obtained from an Anderson impurity model with hybridization
function
\begin{equation}
\Delta (i \omega) = \sum_{k} \frac {{V_k}^{2}}{i \omega-\epsilon_{k}}
\end{equation}
provided that $\Delta (i \omega)$ obeys the self-consistency condition:
\begin{equation}
{t^{2}} G(i \omega) {[\Delta]} = \Delta (i \omega).
\label{mf}
\end{equation}
Here,
 $G(i \omega) [\Delta]$ is the Green's function of the SIAM
(single impurity Anderson Model)
\def\sumks{\sum_{K \sigma }}
\begin{equation}
\sum_{K \sigma } \epsilon_{K} c^{+}_{k \sigma} c_{k \sigma} + \sumks
V_{k}(c^{+}_{k \sigma}f_{\sigma}  + {f_\sigma}^{+} c_{k \sigma}) + U f^{+}_{\uparrow}
f_{\uparrow} f^{+}_{\downarrow} f_{\downarrow} =
H_{SIAM}.
\label{siam}
\end{equation}
viewed as a functional of the hybridization function $\Delta (i \omega)$ which is the Hilbert
transform of $\sum_k {V_k}^2 \delta(\omega-\epsilon_k)$.
\def\deps{d \epsilon D( \epsilon)}  
Equations \ref{total} and \ref{kinetic1} express the total energy
of the lattice model
in terms of the { \it local } Green's function of the problem.
We can therefore express the total energy in terms of the local
spectral function $\rho(\omega)= {-1 \over \pi} Im G(i \omega= \omega+ i \delta)$
using the spectral representation 
$G(i \omega_n) = \int d\omega { {\rho(\omega)} \over {(i \omega_n - \omega)} } $
$ =  \int { { \deps}   \over {(i \omega_n + \mu  - \epsilon  - \Sigma ( i \omega ) )} } $
with $D(\epsilon)$ the semicircular lattice density of states:

\def\ener{
2 t^2 \int \int d \omega_1
d \omega_2 f(\omega_1) {\rho(\omega_1) \rho(\omega_2) \over{\omega_1 - \omega_2}}
}
\begin{equation}
E= \int f(\omega) (\omega +\mu)  \rho (\omega) + \ener
\label{naive}
\end{equation}
 We  will work in the grand canonical ensemble
with the chemical potential chosen to be equal to $\mu=U/2$.
$f(\omega) $ is the Fermi function.
\section{Evolution of the Spectral Function at Zero Temperature}
\label{zrk}
In this section, we describe  the qualitative features of the 
evolution of the spectral function, as  a function of interaction
strength  $U/t $
which is obtained by
solving the mean field equation \ref{mf} at zero temperature.
These  features were discovered in an  
IPT \cite{zrk} study  by Zhang, Rozenberg and Kotliar.

We start   at large U  with a paramagnetic  insulating
solution with  a gap $\Delta (U) $.
When U is reduced below a critical value of U, denoted by $U_{c2}$ , 
(with $\Delta_g \equiv \Delta(U_{c2})\neq 0$)  the 
paramagnetic Mott insulator
becomes unstable against the formation of a metallic resonance at
zero frequency.

The   mathematical
description of the ZRK scenario  of the evolution
of the spectral function when $U_{c2}$ is approached
from below 
is the following: 
\def\gaphalf {{{\Delta_g} \over 2}}
\begin{itemize}
\item[3.1]
$\mbox{Im}\,G(\omega,U) \neq 0$ for all $| \omega | < {{\Delta_g} \over 2}$ and  for all $U <
U_{c_2}$ (finite spectral density everywhere in the metallic phase). 
\item[3.2]
$\lim_{U \rightarrow {U_{c_2}^{-}}} \mbox{Im}\, G(\omega,U) = 0$ for fixed
$\omega$ such that $0 < |  \omega | < \gaphalf $ (Existence of a finite gap
at the Mott transition point).
\end{itemize}
We now discuss more delicate issues, in which the frequency approaches
zero while at the same time,  $U$ approaches
the critical value $U_{c_2}$. More specifically, we define 
\def\ww{\tilde w}
$\ww\equiv\frac{U_{c_2}-U}{U_{c_2}}$ and take the limits $\ww\rightarrow0$
and $\omega\rightarrow0$ such that $x\equiv \omega/ \ww$ is fixed. 
This limit defines the scaling functions which were computed in
ref \cite{moeller1}.
\begin{itemize}
\item[3.3]
  
${\lim_{U \rightarrow U_{c2}^{-} }}\mbox{Im}\, G(\ww x,U)\ne 0$,
i.e.   there is a finite density
of states at the Fermi level all the way up to the transition.
In particular the 
pinning condition which leaves the density of
states at zero frequency unrenormalized is obeyed everywhere in the metallic phase.
\item[3.4]
For a generic value of x
$\lim_{U\rightarrow U_{c2}^{-} }\mbox{Im}\,\Sigma(\ww x,U) $ is finite.

Notice however that  since Fermi-liquid theory
is valid below the Fermi energy in the metallic phase,
for a fixed value of  $U$ below $  U_{c2}$
\begin{equation}
\lim_{\omega \rightarrow 0}
Im \Sigma (\omega , U)=0
\label{fermil}
\end{equation}
\item[3.5]
There exists a $x_{0} \sim O(1)$  such that 
\begin{equation}
\lim_{U\rightarrow
U_{c2}^{-} } Im \Sigma (\sqrt{\ww} x_{0}, U)=\infty
\label{item9}
\end{equation}
\end{itemize}
This incipient divergence  and its significance was recognized in
ref \cite{zrk}: it represents   the precursors of the Hubbard
bands
in the metallic phase.   Its presence is unavoidable, since  spectral
features  resembling the  Hubbard
bands are already
well formed on the metallic side of the transition \cite{ipt}.
The divergence of the self energy occurs outside the Fermi liquid
regime and should not be interpreted in terms of quasiparticle
scattering. It should be understood as 
the precursor 
of the pole  found at zero frequency in the  paramagnetic insulator phase.
This pole indicates that the paramagnetic insulating ground sate, is not
smoothly connected to the non interacting Fermi gas.

It is important to stress, that in  the metallic phase 
the density of states does not vanish  for energies less than  
$\Delta_g \over 2 $. This is a simple consequence of the self consistency
condition of the dynamical mean field theory. The statement that
the Mott Hubbard gap  is finite at the Mott transition point,
should be understood in terms of the previously described,
highly non uniform, limiting
procedure.

We stress that the
results discussed above, were derived by non perturbative means.   
The mapping of the Hubbard model in large dimensions, onto the impurity model
can be done using the cavity construction \cite{review} which does
not involve any expansion in U. Furthermore, to reach the conclusions
discussed above, non perturbative treatments of the impurity model
and the self consistency condition are required.  

In  the next section we  mention a perturbative
expansion, the    skeleton expansion,   which expresses
the self-energy as a power series in
\def\Im{I_{\alpha, m}(\omega)}
$U$ and in terms of the fully renormalized Green's function,
\begin{equation}
\label{skeleton}
Im \Sigma (U,\omega) = \sum_{\alpha, m}  I_{\alpha, m}(\omega). 
\end{equation}
Here, $I_{\alpha, m}$ denotes the contribution of a specific Feynman
skeleton graph, labeled $\alpha$ and of order m in the interaction
strength U, to the  imaginary part of the self-energy
evaluated at a frequency $\omega$. 

The convergence  properties of this 
series are not well understood \cite{lange} \cite{hof}.
Since  the Anderson
impurity model with a hybridization
function which is non vanishing at zero frequency, 
has a singlet  ground state which is a smooth function of U, 
it  may converges for very small U. It is  also known
that the series diverges when U
is sufficiently large  and the lattice model
supports a paramagnetic insulating phase.

In the ZRK scenario, since the   graphs of the 
the skeleton series for  
 $   \Im  $ are  evaluated in terms of U and   G
which  has a very small  spectral weight at low frequencies

\begin{equation}
 {\lim}_{U \rightarrow U_{c_2}} I_{\alpha,m}(\omega)=0
\label{item11}
\end{equation}
 for all
$  \gaphalf > | \omega |  > 0$, 

But  Eqs. \ref{item9},  \ref{fermil}
imply that  the  function
which the skeleton expansion   represents in some form,
behaves very differently  in various frequency reanges.  
So  even if the skeleton  expansion
converges pointwise  in the open interval  $ (0 , U_{c2}) $  
the convergence in this interval cannot be uniform.
Finally, we notice that exactly at the point $U_{c2}$,  the
quasiparticle peak has zero weight. The system  is in the paramagnetic
insulating phase  where the skeleton series
is known diverge.    

The lack of  
uniformity in the frequency domain,  is the mathematical manifestation of the
collapse of  the Fermi energy, as we approach the
transition. Below that scale  a  power  series 
in the interaction has to be well-behaved because
at low frequencies the system resembles a  correlated metal, 
which is smoothly connected to the non interacting system
by  Fermi liquid theorems. 
At high frequencies, the system resembles a paramagnetic insulator,
which has a doubly degenerate  ground state at each site.
For such a system skeleton perturbation theory is known to diverge,
because a doublet cannot be smoothly connected to a singlet
ground state.

\section{Critiques  of the  ZRK scenario}
\label{criticism}
Some of the  findings in  the  ZRK paper  described in the previous section 
were expected.  For example, the gradual narrowing of the
resonance as the Mott transition is approached, is
the  result of the Brinkman Rice mass enhancement
\cite{rice}.  
Other aspects of the ZRK  scenario,  however,  were
new  counterintuitive  and surprising. The instability
of  a  Mott insulator {\it with a finite gap }, towards
metalization was unexpected (previously, such an instability was
only expected to take place when the gap was infinitesimal).
Also the incipient divergence of the self energy, at a relatively
high energy scale
$\sqrt{\ww} D$
had not appeared in earlier  slave boson studies.

The  alternative scenario for the Mott transition in frustrated systems, 
is a bicontinuous one (i.e. continuous from the metallic and the insulating
side)   In this scenario
the  gap closes 
from the insulating side   at the same critical value
of the interaction
at which  the resonance vanishes upon  approaching the transition
from the metallic side.
This bicontinuous 
scenario was  shown to occur within the slave boson formulation
of Kotliar and Ruckenstein \cite{kr}
after including Gaussian fluctuations 
on top of the mean field theory \cite{fresard} \cite{castellani}
and within a large N model of the metal
to charge transfer insulator \cite{castellani}.
The natural extension of this scenario to finite temperatures
gives a smooth crossover between a metal
 \label{bicon}
and an insulator, excluding
a first order phase transition between a metallic and an insulating
phase but other extensions are possible \cite{fresard2}.

Several  authors  raised questions about   the physical meaning
and the internal consistency of the ZRK scenario and raised the
possibility    \cite{gep}
\cite{logan} \cite{nozieres}
 \cite{kehrein}
\cite{schlipf} 
 that  
the alternative, bicontinuos  scenario might be realized in  the frustrated
large d Hubbard model with the semicircular density of states.

Nozi\`eres \cite{nozieres}
expressed the surprising nature of the  evolution
of the spectral function at  zero temperatue 
in two different ways. First an  energetic argument:
if one is near the Mott transition, transfer of a   small amount
of spectral
weight 
$\epsilon$ 
from high to low energy   across a Mott Hubbard gap
$ \Delta_g$ costs $\epsilon \Delta_g$ while
the kinetic energy gain is only of order $\epsilon ^2 D $. For the
transition to take place,  these two terms  have to balance
(i.e.  $ \Delta_g$ has to be of order $\epsilon D$)  thus ruling out
the possibility of a   preformed Mott Hubbard gap.

The second argument makes use of
Nozi\`eres exhaustion principle
( which we interpret in the present context as saying that
the Kondo quenching of a spin in a  dilute bath of conduction electrons
does not result in a substantial energy gain).
Near the Mott transition,  the ZRK spectral
function  has very few carriers,
while there are many  spins.
The  system is 
in the exhaustion regime, where very  little Kondo energy can be
gained, and  is therefore unlikely to be energetically stable.
Full Kondo compensation, which is necessary to form
the resonance, and the Fermi liquid state  is unlikely.

Recently Kehrein  
argued  \cite{kehrein},
that under the assumption that the sekelton expansion
converges pointwise in the   open
interval $(0,U_{c2})$, 
the scenario
presented in the previous section is inconsistent.
From equation    
\ref{item11} 
he concludes
\begin{equation}
\sum_{\alpha, m}{\lim_{U \rightarrow  U_{c2}}} I_{\alpha, m}
(\omega = x_0 \sqrt{\tilde w})= 0 
\end{equation}
He then  interchanges  the order of the  infinte summation
and  the limiting procedure. Using equation 
\ref{skeleton}  
he obtains

$
\lim_{U\rightarrow
U_{c2}^{-} } \Sigma (\sqrt{\ww} x_{0}, U)=0 $  contradicting eq.
\ref{item9}.

This exchange of limits, is not allowed, even when
the  skeleton expansion converges pointwise for all frequencies,
and even if the total  spectral weight of each graph can
be  bounded  uniformly as  is stated in footnote 14 of ref
\cite{kehrein}.
 
The exchange  of limits is  mathematically
justified only when the  series
\ref{skeleton}
converges {\it uniformly } .
The skeleton expansion cannot converge {\it  uniformly }
as one approaches the Mott transition since it  
diverges at $U_{c2}$, which in the ZRK picture,
is in the paramagnetic insulating phase.

Recently
E Lange \cite{lange} analyzed  the soluble strong coupling
limit  of
an Anderson  impurity for arbitrary U.
He showed explicitly that while the skeleton expansion
may fail to converge, particularly, 
at high frequencies,  an IPT-like  treatment, 
gives the exact answer for this model.

The  rest  
of this paper addresses  the questions  of 
Nozi\`eres.
The  goal is to    explain why at some point it becomes
energetically favorable to metallize a paramagnetic
insulator while it still has  a finite gap. We also clarify
how the energetics of lattice models can be estimated
using impurity models.
For these purposes, we develop a framework 
based on a Landau like functional to connect
lattice and impurity  physics in the next section.

\section {Landau Functional}
\label{functional}

To address the questions about energetics and to  illuminate
the analogies and differences between 
the dynamical mean field theory and the more standard
Landau theory of phase transitions it is useful to introduce 
the free energy functional \cite{vlad}:
\begin{equation}
F_{LG}{[\Delta]} = - T  \sum _{\omega} \frac{\Delta(i \omega)^{2}}{t^{2}} + F_{i
mp}
{[\Delta]}
\label{landau}
\end{equation}
where $F_{imp}$ is the free energy of the impurity model defined  by
$\ref{siam}$ which  can be represented as a functional integral:

\def\sumo{\sum_{\omega,\sigma}}
\begin{equation}
\label{func}
e^{-\beta F_{imp}} = \int df ^{+} df e ^{- L_{loc}[f^{+},f] - 
{\sumo}_{\sigma} f^{+}_{\sigma}(i\omega) \Delta (i\omega)
{f_{\sigma}}(i\omega)}
\end{equation}

Here, $L_{loc}$ is the action  of a local f level with the
hybridization set to zero. 

One should regard equation
 \ref{landau} as a
  Landau Ginzburg Functional of the ``Metallic
Order Parameter'' $\Delta(i \omega)$.
At zero temperature,
the function $\Delta(i \omega) $ is non zero in both phases, but
it has very different low frequency behavior in the metallic and in
the insulating phase.

Differentiating  the free energy
with respect to 
$\Delta(i \omega)$
we obtain the mean field equation \ref{mf}. The derivative
of the first term is 
$- 2 \frac{\Delta(i \omega)T}{t^{2}}$.
Combining it
with the results of  differentiating  the impurity free energy in 
\ref{func}     
\begin{equation}
\frac{\delta  F_{imp}}{\delta  \Delta(i\omega)} =
 T \Sigma_{\sigma}
<{f_\sigma}^{+} (i\omega) f_\sigma (i\omega)>
 =  2 T  G(i\omega)[\Delta] 
\label{gimp}
\end{equation}
we obtaine eq. \ref{mf}.

We will show below,  by means of 
an explicit  calculation, that 
$F_{LG}{[\Delta]}$
evaluated at the saddle point 
\ref{mf}  gives the correct energy of the lattice problem. 
$F_{LG}{[\Delta]}$
allows us   to consider the free energy of arbitrary
hybridization functions 
(i.e. away from self consistent solutions), and to interpolate
between various stationary  points. 

It also  has a nice  physical interpretation by analogy
to the mean field free energy of a spin system
regarded as a function of the Weiss field h:
\begin{equation}
\beta {F_{LG}}[h] = {\beta {{h} ^{2} \over {2J}} -  log[ch[ 2 \beta h]]}
\label{classical}
\end{equation}
The first terms in \ref{classical} and \ref{landau}
are the cost of forming the Weiss fields around a site.
The second terms are   the energies of  a  site (spin in
the classical case, electron in the quantum case)  in the 
the presence of the
the Weiss field around it.

Note that our order parameter $\Delta(i \omega)$ is always non zero but
has
a qualitatively 
different form at low frequencies
on either side of the transition.  
This is analogous to the density in
the  liquid
gas transition.
At zero temperature, there is a qualitative difference between
a metal and an insulator, which appears naturally in this formalism.
The metallic phase is characterized by a hybridization
function which is nonzero at low frequencies, as opposed to insulating phases
for which the hybridization function  vanishes at low energy \cite{caveat3}.

We  stress the difference between the Landau functional
\ref{landau} and expression \ref{naive}.
Both  \ref{naive}  and
\ref{landau}   give the correct
energy if one evaluates them  at the {\it exact local spectral function}
or the
{\it exact hybridization}  function respectively.
However,   \ref{landau} is a stationary functional,
i.e. upon differentiation it produces
the correct dynamical mean field equations.
On the other hand,  \ref{naive} is
not stationary 
(its derivative with respect to $\rho$ is nonzero at the physical
spectral density)  and therefore it does
not give the correct mean field equations.
It is not a Landau functional, it  can only  be used to obtain  the ground
state  energy if one knows the correct spectral function
but it cannot be used  to determine the spectral function itself.

We now prove, that  the Landau free energy, evaluated at
the saddle point value of the order parameter,
indeed gives the total energy of the lattice model.
In the process we
make the  correct connection
between   the various contributions to the  energy of
the lattice model and  to the energy of  
the  corresponding impurity model.

The lattice Hamiltonian
contains two terms,  K and U.
When $\Delta (i \omega)$
(or correspondingly the one particle Green's function)
changes,  both
$< K >$ and $< U >$ change.   $F_{LG}[\Delta]$ has two
terms,  $-T \sum _{\omega} \frac{\Delta (i \omega)^{2}}{t^{2}}$ and $F_{imp}
[\Delta]$,  but  it would be incorrect to associate $F_{imp} [\Delta]$
directly with the kinetic energy of the lattice model. 
At a stationary point and
at T = 0, the correct connection is:
\begin{equation}
F_{imp}[\Delta] = \frac{3}{2} < K > + < U >
\label{impurity}
\end{equation}
Notice that since we
work at zero temperature,   the free energy and the energy
are identical.
The impurity energy is composed of three parts, the local
correlation $< U >$ , the hybridization energy
$2 \sum_{k  \sigma} V_k <f_{\sigma} ^{\dagger}c_{\sigma k}>$
which is given by  $4  T \sum_{\omega} \Delta (i \omega) G ( i \omega)$
(extra factor of 2 comes from spin) and   the change in kinetic
energy of the conduction electrons, which is defined as

$\sum_{k \sigma} \epsilon_k ( <c^{\dagger}_{\sigma k}  c_{\sigma
k}{>_{\Delta} - <c^{\dagger}_{\sigma k}  c_{\sigma k}>_{\Delta =0}})$

and is therefore given by
\begin{displaymath}
\nonumber
2T \sum_{\omega,k}{{{V_k}^2 \epsilon_k} \over { (i \omega -
\epsilon_k)^2}} G(i \omega)
\end{displaymath}
\begin{displaymath}
 =  -2T \sum_{\omega} \Delta(i\omega) G (i\omega) + 2T \sum_{\omega}
\frac{1}{2} [\frac{-d}{di\omega} (\Delta(i\omega)G(i\omega)] i\omega
 =  -T \sum_{\omega} \Delta (i \omega) G(i \omega)
\end{displaymath}
Combining the three terms,  we arrive at equation \ref{impurity}.
The first term in
the functional in eq    \ref{landau}
is given by
\begin{equation}
- T \sum_{\omega} \frac{\Delta (i \omega )^{2}}{t^{2}} =
\frac{-1}{2}  \langle  K  \rangle    >0
\label{kinetic2}
\end{equation}
The positivity of this term is essential
if we want to interpret it as the cost of forming the Weiss field.
Clearly once the Weiss field is formed the impurity energy is
negative.
(Notice that we are in the half filled situation where
 $\Delta( i \omega) $is purely {\it imaginary}).

Notice that
while local quantities have a straightforward
interpretation in terms of the impurity model, nonlocal quantities such as
the kinetic energy require more care.
This is why we stated very clearly how the self-consistent
local impurity approximation embeds the impurity model in a medium
and gives well defined relations between the kinetic and potential
energy of the lattice model, and the energy of the Anderson impurity model,
 eqs  \ref{kinetic1}, \ref{kinetic2}   \ref{landau}, and 
 \ref{impurity} are exact in the limit of large dimensions.

The previous discussion highlights the  self consistent
character of the Local Impurity Self-Consistent Approximation,
whereby the energy
is composed of two terms, cost of the Weiss field and   the energy
gain of the impurity in the presence of the Weiss field.
This is very different from the reasoning within the local impurity
approximation,   which would equate the energy of the
lattice to the number of sites times the impurity energy.

\def\dl{\delta {\Delta_L} (i \omega)  }
\def\ddl1{\delta {\Delta_L} ( \omega)  }

Nozi\`eres \cite{logan} \cite{nozieres} reduced 
the validity of the semicontinuos or the continous 
scenario to the following question: does one gain Kondo energy once per spin or once per electron
in the resonance (described by the  small fraction of  electrons contained
in the low
energy part of the bath)? .

To answer this question  an unambiguous expression of the 
energy of the lattice models  in terms of  the energy of  impurity
models  in the {\it putative  trial states}
is required. 
Expression \ref{landau},  unambiguously
describes the energy per lattice site, i.e.
the energy per spin, in a  self consistent local impurity framework
substantiating the semicontinuous scenario.

We can now describe in a precise manner the energetics
near the critical points $U_{c1}$ and $U_{c2}$ .
Their existence highlights an  essential difference
between the classical and the
quantum Landau Ginzburg functionals.
While in the classical case the Landau functional is
an {\it analytic } function of the magnetization and of the
Weiss field which has a straightforward power series expansion,
this is not the case in the quantum case.

Free energies obtained by  "integrating out" massless
degrees of freedom  ( fermions with a Fermi surface) are
non analytic. Here we will deal with the manifestation
of this problem in infinite dimensions.
We  note however that
while this problem is most extreme in infinite dimensions,
some manifestation of this problem in weaker form may
be relevant to
 the  treatment of quantum
phase transitions  in finite dimensions.

An important  lesson to be learned from this analysis is that
the quantum Landau Ginzburg functional is non analytic
in the field $\Delta$.
The source of non analyticity  can be traced to the presence of a Fermi surface,
and  in infinite dimensions can be given a sharp formulation
following Nozi\`eres' analysis of the Kondo model.

\section{Projective Self Consistent Analysis}
\label{projectivesec}

In the next sections we  address  a number of
questions concerning the stability of the paramagnetic
insulating state
 $\Delta_{o}(\omega) $.
\def\d0{ \Delta_{o}(\omega)}
When we evaluate the free energy functional in
the perturbed state $\d0+\ddl1$  (i)
do we gain or lose energy ? (ii) and  how much? (iii) What is the 
physical origin of
the energy gain ?

For this purpose we analyze the  low energy
behavior of   the  Landau  functional, using 
the projective self consistent method, 
a  technique  designed 
to isolate the low energy  physics  of impurity models
in 
a bath which is self consistently determined (as in the solution of
lattice models in the limit of  large lattice coordination)

The goal of this section is to set up the notation associated
with   this reduction to low energies. 
The results quoted here  will be used in the next few sections
to  isolate and discuss
the singular dependence of
the  Landau free energy on
the   metallic order parameter.
The projective  self consistent method  was developed
in references \cite{moeller1}   \cite{moeller2} and \cite{thesis} 
where it  was used to calculate the value of $U_{c2}$  
and the  scaling functions at the Mott transition point.

We start with   
a Mott paramagnetic insulating solution,
$\Delta_{o}(\omega) $, i.e. a solution to the mean field equations
\ref{mf} with a finite gap and we  add to it a perturbation
\def\uno {\delta {\Delta_L} (i \omega) = }
\def\tres{{i \omega - \epsilon_k}}
\def\dos{{{{\Vkl}}^2}}
 $   \uno  \sum_{{\varepsilon_{k}}  }  {\dos \over \tres} $
localized
in the low energy region
(i.e.  the variables $ \varepsilon_{k}$ are much smaller in
absolute value than the Mott Hubbard gap).

Notice that since $\Delta_0$ is the exact insulating solution,
(i.e.   a stationary point of the functional \ref{landau} )
the expansion of the energy in the small addition
$  \delta\Delta_L (\omega)  $ starts with quadratic terms
We will therefore only write the terms which are quadratic
in  $\dl$, and ignore the linear variations which  have
to cancel in the final answer \cite{caveat}.

Furthermore we take $ \delta \Delta_{L}(\omega) $
as  entirely concentrated in the region
of low energies, since changes at high energies will be shown to be unimportant
(see discussion of this point on page   \pageref{16th}).
Using a  Schrieffer Wolff\cite{wolff} transformation on the Hamiltonian,  we can
reduce
\ref{siam}  (to order ${\Vkl}^2$) to a Kondo like Hamiltonian
\begin{eqnarray}
H_{K} (\delta \Delta_{L}) = E_{ins}+
 J_s(U,t)
 \sum_{kk^{\prime}} \Vkl c^{\dagger }_{k}{\vec {\sigma}}
c_{k^{\prime}}
\Vklp  \vec{S}
 + \sum_{k} \epsilon_{k}
c^{+}_{k \sigma} c_{k \sigma} 
\label{kondo0}
\end{eqnarray}
\def\cl{c_{L \sigma}}
$ E_{ins}$ is the energy of the Mott insulator, the subscript k, runs
over the low energy conduction electrons.
\def\ss{\sigma \sigma^{\prime}}
We  introduced the local conduction electron:
\begin{equation}
\cl = 
\sum_{k}   2 { {\Vkl c_{k \sigma} } \over   { D  \sqrt{Z}}}
\label{locel}
\end{equation}
and ${\vec S}$, 
the local renormalized  impurity spin $ {S^a}={1 \over 2} {\sigma^a}_{\ss}
X_{\ss}$
acting on the low energy spin
degrees of freedom.
${|\downarrow\rangle}_{a}$,  
${|\uparrow\rangle}_{a}$.
These are defined formally as 
 ground states
${|\downarrow\rangle}_{a}$,  
${|\uparrow\rangle}_{a}$, 
 of the Anderson impurity model
 with hybridization function $\Delta_0$
 \cite{moeller1}.
X are Hubbard operators, acting on those spin degrees of freedom.

$J_{s}(U,t)$ is a
monotonically decreasing function of $U/t$
which depends explicitly
on the insulating solution $\Delta_0$. An explicit expression
is given by:
$ J_{s} \equiv   {\ _a \langle \uparrow|} f_\downarrow
\frac{1}{H_{a}-E_{g}^{a}}f^{\dagger}_{\uparrow}{|\downarrow\rangle}_{a}$
\cite{moeller1}.

Since we intend to compute the energy change to order
$\dl^2$, we need in principle the effective Hamiltonian
obtained by  the Schrieffer Wolff\cite{wolff}  transformation to
order ${\Vkl}^4$.
It has the form:

\begin{eqnarray}
 H_{low}^{(3)} = &
\frac{D}{2} J_{1}^{(3)} \vec S\cdot \vec s_{\bar\epsilon L}
+\frac{D J_{2}^{(3)}}{8}\sum_\sigma (c^\dagger_{L\sigma}
c_{\bar\epsilon L\sigma}+ c^\dagger_{\bar\epsilon L\sigma} c_{L\sigma})
 \\
 &+{ \frac{D}{2}} {J_{3}}^{(3)} \vec S\cdot \vec s_{L}
+{\frac{D}{16}}
 {J_4}^{ (3)}
 (n_{L\ua} - {1 \over 2})( n_{L\da} -{1 \over 2})
\label{Heff_K2}
\end{eqnarray}

with
$c_{\epsilon L\sigma} \equiv
2 \sum\limits_{k} {\tilde{V}_k} \epsilon_k c_{k\sigma} $ and
 $\vec s_{\bar\epsilon} \equiv \frac{1}{4}(
c^\dagger_{\bar\epsilon \alpha} \vec \sigma_{\alpha\beta} c_{\beta}
+c^\dagger_{\alpha} \vec \sigma_{\alpha\beta} c_{\bar\epsilon\beta})$.
The coefficients $J^{(3)}$, have been evaluated numerically
\cite{thesis} but they are not necessary  for our purposes since
the expectation
value of  the fourth order Hamiltonian
\ref{Heff_K2}
in the ground state is zero, hence  contributions
to the energy to order $\dl^4$ only arise from a second order
computation using eq.  \ref{kondo0}.

It is also instructive to record the explicit form of the
low energy part of the  Anderson model  f operator
after the canonical transformation is performed:
\def\gam{{J_{s} \over 2}}
\begin{equation}
F_{\sigma} = 
 -  \gam D \sqrt{Z}   \sum_{ \sigma' } {S^a} {\sigma^a}_{\sigma, \sigma'}  c_{L \sigma'} 
\label{fop}
\end{equation}

Z is a very important quantity, 
the total low energy spectral
weight of the trial state:

\def\normal{ { 4 \over {D^2} } }
\begin{equation}
Z =
\normal
 \int
 \delta \Delta_{L}(\omega^{\prime})d\omega^{\prime} = 
 \sum_{k} 4  {{{\Vkl}^2} \over{ D^2}}
\end{equation}
It is 
 proportional
to the quasiparticle residue of Fermi liquid theory.
The proportionality constant between the low energy spectral
weight and the quasiparticle residue was evaluated in ref
\cite{moeller1}.

We now  isolate the dependence of
the  Landau functional on $\delta \Delta_{L}(i \omega)$, from eq. \ref{landau},
namely, 
\begin{equation}
E_{L} (\delta \Delta_{L}) =   \frac{-T }{t^{2}} \sum_{\omega}  \delta
\Delta_{L}( i \omega)^{2}  + E_{K} (\delta \Delta_{L})
\label{projective}
\end{equation}
$E_K$ is the usual
expression for the energy of a Kondo impurity model
\begin{equation}
{E_{K}}(\delta \Delta_{L}) = < H_{K} > [{{\Vkl}}] - < H_{K} >[{{\Vkl}}=0]
\label{energy} 
\end{equation}

While one
can make qualitative arguments about
transfer of spectral weight by regarding $E_L (\delta \Delta_L ) $
as a {\it function }of $Z$
one should always keep in mind that 
$E_{L}$ is really a {\it functional} of the whole hybridization
function $\Delta$  ( or, if we restrict ourselves to low energy
variations,  of $\delta \Delta_L$).
In  fact, we shall show
in the following sections  that the free energy functional
(\ref{projective})
takes  very different  values    for hybridization functions
having the same value of Z 
but different aspect ratios or  shapes.
The physical reason underlying this   is
the different behavior of the  Kondo energy 
depending
on whether the model \ref{kondo0} is in the weak coupling
or the strong coupling limit.
This will have significant implications for the Mott transition
in the limit of infinite dimensions.

\section{Energetics near $U_{c1}$}
\label{uc1sec}

In this section we describe the energetics of making a "very small "
perturbation (very small is defined by the requirement that   
the  corresponding
effective Kondo problem  \ref{kondo0} is weakly coupled) to
the Mott insulating solution.
The goal is to determine the location of
$U_{c1}$,   the point
where the insulator ceases to exist because it becomes
linearly unstable against such a small  perturbation.
\cite{moeller2}.  

The  cost of modifying  the Weiss field (first term
in
\ref{projective} ),  
is computed
by  inserting explicitly the expression $\delta \Delta_{L}(i\omega_{n}) =
\sum_{k} {{{\Vkl}}^2 \over { i\omega_{n}-\varepsilon_{k}}}$,   doing
the Matsubara sums,  and taking the zero temperature limit, and is given by:

\def\vkl{V^{L}_{k}}
\def\vkpl{V^{L}_{k^\prime}}
\begin{equation} 
{2 \over {t^2}} \sum_{{\varepsilon_{k'}} > o,{\varepsilon_k} < 0 }
{\num1 \over
 \den}
\label{cost}
\end{equation} 

Next we evaluate eq. \ref{energy}  
in second order perturbation theory   to compute the  
energy gain

\begin{equation}
E_K (\delta \Delta_L ) = - \sum_{\varepsilon_{ k^{\prime}}  >0,\varepsilon_{k} < 0}
{{(J_{s} \vkl \vkpl)}^2  { <\phi_0|{\vec S} . c^{+}_{k\sigma} \vec{\sigma} c_{k' \sigma}
c^{+}_{k'\sigma} \vec{\sigma} c_{k \sigma} {\vec S} |\phi_0  >} \over
{(\varepsilon_{k'} - \varepsilon_{k}})} 
\label{2pt}
\end{equation}
 $ |\Phi_{o} >$ is a Fermi sea, i.e. a ground state of
 $\sum_{k}
\varepsilon_{k}{ c^{+}}_{k\sigma} c_{k\sigma}$.
Evaluating \ref{2pt} we get
$E_K (\delta \Delta_L )= {J_{s}}^2
{\frac{3}{2}}
\sum_{\varepsilon_{k'} >o \varepsilon_{k} < 0} { {\Vkl}^{2}  {\Vklp}^2
\over {(\varepsilon_{k^{\prime}} - \varepsilon_{k})}}$
Therefore, the energy gained by a {\it weakly coupled Kondo impurity}
embedded in the perturbed  Weiss field 
is 
\begin{equation}
E_K (\delta \Delta_L )=
-T[{\sum_\omega}  \delta{ \Delta_{L}(i \omega)^2}  ] {3 \over 4 } {{J_s} ^2}
\label{gain}
\end{equation}
Balancing cost and gain then gives the condition for
$U_{c1}$, namely
\def\js2{{J^{2}_{s}}(U,t)}
$\frac{1}{t^{2}} = { \frac{3}{4}}\js2 $.
Recalling that $ t = \frac{D}{2}$  we have rederived the equation
that determines $U_{c1}$  first obtained in ref \cite{moeller2}:
 
\begin{equation}
\frac{1}{D^{2}} = \js2 \frac{3}{16}
\label{uc1}
\end{equation}

Since  the closure of the gap, necessarily implies instability
towards metalization, the  U that solves eq \ref{uc1} is an upper bound to
the value of U at which  the gap  closes \cite{caveat2}.

An advantage of the Landau formulation is that the trial states
$\dl$ do not have to be normalized to unity, i.e. the condition
$1=\sum_{\omega}  {{\Delta (i \omega)}  \over t^2} \ee$ does not
have to be satisfied. However, if for esthetic reasons one wants
to exhibit trial states which do satisfy a proper normalization,
this can be easily done without altering significantly the energy
of the state by 
 adding  a small perturbation 
$\delta \Delta_H$ centered at high energies  \label{16th}
near the gap edge.
Since $\Delta_0$ was stationary, the cost associated with this addition is
of the form
\def\weight
{\sum_{\varepsilon_{k}}{\Vkl}^2}
\begin{equation}
\Delta E \approx
{({a_1}\weight+ {a_2}{V_{H}}^2) {V_H}^2
\over D} 
\end{equation}
\label{normalize}
with  $a_1$ and $a_2$  numbers of order unity.
Since 
${V_{H}}^2 $  is of order Z, but the denominator D is much larger than
the corresponding denominator in equations
 \ref{cost} and
 \ref{2pt} it is clear that this
term does  not contribute  substantially to the energy balance and the 
condition  for the insulator to become unstable is  still controlled by the
equation involving low energy denominators (\ref{cost}, \ref{2pt}) .

The  first argument of Logan and Nozi\`eres, which stated that
we can only gain energy from the metalization process when
the gap has closed    is   in complete
agreement with our detailed calculations,    but  is  only valid
for trial states, relevant to the determination of
$U_{c1}$.
Under the assumption that the trial state
is in the weak
coupling regime,  we can
regard the changes in  the spectral function as small continuous
deformations of the insulating solution.
This kind of perturbations destabilize the  insulating solution
only when the gap closes.

We can now address the second point raised
by Logan and Nozi\`eres.
What is going on from
the point of view of  Nozi\`eres exhaustion principle:
a few electrons cannot gain a great deal of Kondo energy when
they have to  screen a lattice of  spins.
We see that this argument is qualitatively correct
when applied to the dynamical mean field theory
provided that the conduction band is
weakly coupled to the spin. 
Indeed, if  we regard the variables $\epsilon_k$ in
eqs \ref{cost} and  \ref{gain} to be of order D, (and not of order ZD),
the changes in energy are indeed of order $Z^2 D$

More generally, we note that the partial screening of magnetic moments,
is a very common situation 
in  the full solution of the non linear
dynamical mean field equations \ref{mf}.
It occurs  in  the correlated
regime at {\it finite temperatures}
when the condition  $\Delta (i 0^+) = - i t$
is strongly violated.

\section{Energetics near  $U_{c2}$}
\label{uc2sec}
A most interesting aspect 
of the large d solution is   the instability of the
Mott insulating state with a finite gap to the formation
of a narrow metallic band at the Fermi energy.
This takes place at a critical value $U_{c2} > U_{c1}$, as we now show.
 
We are interested in the energetics of this problem which
we shall consider from the point of view of the free energy
\ref{naive} and \ref{landau}.
We first consider the energetics of the problem by inserting a spectral
function $\rho(\omega)= \rho_0(\omega)+\delta\rho(\omega)$   in the functional
\ref{naive}. $\rho_0$ describes the Mott insulating state, and $\delta \rho$ is comprised
of two pieces, one removing spectral weight  Z 
from the gap edge at $-\Delta_g$ and one that
adds it at zero energy. 

Since the functional \ref{naive} is quadratic, the evaluation of the energy difference
between this state (our candidate for metallic state) and the Mott insulating state
$\rho_0$ is straightforward.  It consist of a linear part  in $\delta \rho$, $\delta E_1$ and a quadratic part in $\delta \rho$, $\delta E_2$
. These are given  respectively by:
\def\om1{\omega_1} 
\begin{equation}
\delta {E_1}=\int d \om1 \delta\rho(\om1) [\om1 f(\om1)+ {2 t^2} \int d \omega_2
{\rho_0} (\omega_2){{f(\omega_1) -f(\omega_2)} \over {\omega_1 - \omega_2}}]
\end{equation}
\begin{equation}
\delta {E_2}=
2 t^2 \int \int d \omega_1 
d \omega_2 f(\omega_1) {\delta\rho(\omega_1) \delta\rho(\omega_2) \over{\omega_1 - \omega_2}}
\end{equation}

It is now straightforward to evaluate $\delta E_1$, which as expected turns out to be
positive representing a {\it cost} of energy  and  given   by
\begin{equation}
\delta E_1 =Z [\Delta_g + 2 t^2 {\int_{0}}^{\infty} d\epsilon {\rho_0}(\epsilon)
 { -\Delta_g  \over {[\epsilon + {\Delta_g}]\epsilon }}]
\end{equation}

\def\cll{{c_L}}

We can now see the breakdown  of the exhaustion arguments
 when applied to the impurity model {\it relevant to $ U_{c2}$ } . If 
$\delta \rho$ were indeed a "small" (in the sense of the previous section) perturbation,
it would then be true that the contribution from 
$\delta E_2$ is negative but small, of the order $Z^2$ and that to obtain a net gain
from $\delta E_1 +\delta E_2$ one would require the gap to be of order Z rather than finite.
However the perturbation that we are considering (which is a perfectly legitimate trial
state, for the large d Hubbard model) is {\it not small}. The low energy part ($\delta\rho =
\delta{\rho_L} + \delta {\rho_H}$) has height one and is described by the scaling
function $\delta \rho_L (\omega) = {1 \over D} \phi ({\omega \over{ Z D}})$.  
The contribution from $\delta E_2$ is {\it of order Z} and is given by
\begin{equation}
\delta {E_2}=
 Z  t \int \int d \omega_1 
d \omega_2 f(\omega_1) {\phi(\omega_1) \phi(\omega_2) \over{\omega_1 - \omega_2}}
\end{equation}
Now it is completely clear that the gain in kinetic energy,   $\delta E_2$ is of the
same order as the positive contribution  $\delta E_1$ so an instability to the metallic
state when the insulator has a finite gap can take place.
To see that it {\it does take place} we need to resort to the variational functional
\ref{landau}.

To demonstrate this point we
take an  insulating
solution with a finite gap with U slightly bigger than
 $U_{c1}$.  According to the  analysis of the previous  section this
is stable against a perturbation  $\delta \Delta_L$ which
is in the weak coupling regime. 
On the other hand   this state can be 
unstable against a perturbation  $\delta \Delta_L$  
described by a Kondo model in an intermediate coupling regime.
In  this section we will concern ourselves with   a perturbation 
$\delta \Delta_{L} (\omega)$ 
{\it  with height of order unity  }.

We will  take the  perturbation with the scaling form
\begin{equation}
\delta \Delta_{L}(\omega) = D {\phi_1} ( {\omega \over {Z D}})
\label{scalingbath}
\end{equation}
with  $ {\phi_1}(0)$ of order unity.  
It is instructive, following  reference \cite{moeller1}, to write 
the low energy Kondo Hamiltonian corresponding to such a perturbation
in a way that makes completely clear that the condition of unit height
of  $\delta \Delta_L (0) $ corresponds to
an  effective impurity model in the intermediate coupling regime (Kondo
energy comparable to electron kinetic energy).
\def\ss{\sum _{  \sigma}}
\begin{equation}
H_{K} = Z  J_{s}(U,t) \cll^{+} \vec{\sigma} \cll
\vec{S} + \ss \sum _{   { - Z D < \epsilon_{k} < Z D}} \epsilon_{k} c^{+}_{k \sigma}
 c_{k \sigma}
\label{kondo}
\end{equation}
It describes a band of narrow electrons with bandwidth ZD
interacting with an orbital
the local electron of equation \ref{locel}
which is
properly normalized
$(\{ c^{\dagger }_{L} c_{L} \}= 1)$, 
via a Kondo exchange  $Z J_{s}(U,t)$.
The Kondo
problem in \ref{kondo} has a coupling 
$J_{eff}= 
  Z J_{s}(\frac{U}{D})$
and an effective  bandwidth
$ D_{eff}=  
 Z D$.
\def\locspin{{\vec S_L}}
As a result $\frac{J_{eff}}{D_{eff}}$
is independent of Z, so the Kondo impurity is in the intermediate
coupling regime.  Therefore  $\langle \locspin {\vec S} \rangle  < 0 $  where
we defined the local conduction electron spin operator by
\begin{equation}
\locspin=  { 1\over 2} {c^\dagger}_L {\vec \sigma }
c_L
\label{locspin}
\end{equation} 
Notice that, while  the value of the expectation value of the
scalar product of the impurity spin and the local spin operator
 depends on the  full form of the scaling function and
 has only been  calculated numerically  \cite{moeller1},
the fact  that it is non zero depends only on the intermediate
coupling nature of the associated impurity model in eq \ref{kondo}.

We now 
 insert the  variation $\Delta_{trial}= \Delta_0+ \delta \Delta_L$,
with  $\delta \Delta_L$
 given by Eq.
\ref{scalingbath}
into the Landau function (\ref{landau}) 
 and  compute  explicitly the quantity 
\def\ff{ { {\partial F_{LG} [\Delta_{trial}]} \over {\partial Z}}} 
\begin{equation}
m \equiv  
\lim_{Z \rightarrow 0} \ff
\end{equation}

If m is positive, there is a finite  net  cost to forming the
resonance, if it is negative then it is energetically 
favorable to metallize.
From  the chain rule, we find
\def\omegan{\omega_n}
\def\kkkk{\delta \Delta_{L} (i \omegan)}
\begin{equation}
m = - 2T \sum_{i\omegan }[ {\kkkk 
\over t^{2}} - G_{L}(i \omegan)]
{d \over {d Z}}   \delta\Delta_{L}(i \omegan)
\label{la}
\end{equation}

We now insert
the spectral representation of  the low energy
part of the  impurity Green's
function $G_L (i \omega) [\delta \Delta_L  ] =
 \int d\epsilon {\rho_L(\epsilon) \over{ (i \omega_n - \epsilon)}}  $ , with
$ \rho_L(\epsilon)=
 {1 \over D}{ \phi}({\epsilon \over
ZD})$, 
into  eq. \ref{la} to obtain
\begin{equation}
m =
4D \int^{0}_{- \infty} dx {\int^\infty_0}  dy \frac{1}{(x-y)}
 [({D \over t})^2  \phi_{1} (x) -
\phi (x)] {\phi_1}^{'}(y) y
\label{m}
\end{equation}
We next assume a trial state such that ${\phi_1}^{'} (y) \equiv 0 $ at small y and
${\phi_1}^{'}(y) = c $ between 
 $W1$ and $W2$. Then
\def\wu{W1}
\def\wt{W2}
eq. \ref{m} simplifies to:
\def\inte{\int^{W1}_{W2} dy } 
\begin{equation}
m \approx 
- 4 D c   \inte   y  {\int^{0}}_{- \infty} dx  \frac{1}{(x-y)} [4 \phi_{1} (x) -
\phi (x)] 
\label{m1}
\end{equation}
For  W1 and W2  large, we thus arrive at: 
\begin{equation}
m \approx 
4 c D (W1-W2) {\int^{0}}_{- \infty} dx   [4 \phi_{1} (x) -
\phi (x)] 
\label{m2}
\end{equation}

$\phi$ is of course a complicated  functional of $\phi_1$.
However,  to estimate a {\it lower} bound to the instability
point
we only need the integral of $\phi$ which we calculated
in   appendix B. Using those results, and the fact that
both $\phi_1$ and $\phi$ are even we
\def\Js{{J_s}}
obtain
the condition for U$_{c2}$ first derived in ref \cite{moeller1}:
\def\sloc{{\vec S_L}}
\begin{equation}
\label{uc2}
1  -  { (D J_s (U)  )^{2} \over 2}[{3 \over 8} - < {\vec S} .\sloc > ] < 0
\end{equation}
Here, $\sloc= {1 \over 2}{c^+}_{L \sigma} {\sigma^a}_{\sigma, \sigma'}  c_{L \sigma'} 
$ is the local conduction electron spin.

The proof of the  instability of an
insulator with a finite gap
results  from comparing  conditions
\ref{uc1} and  
\ref{uc2} which imply  the strict inequality:

\begin{equation}
U_{c1} < U_{c2}
\label{ineq}
\end{equation}

This follows once we recognize
that  while $\langle {\vec S} .\sloc \rangle  $ depends
on $\phi_1$  and can only be computed numerically after solving
a non linear self consistent problem, the fact that it is non
zero and negative is immediate, from the fact that it describes
an impurity in the intermediate coupling regime.
Since  $J_s (U)$ is a monotonically decreasing function of increasing  U,
the solution of eq \ref{uc1} has a lower value than the solution
of equation \ref{uc2}.

Notice that the derivation 
is  free from ad-hoc assumptions.
The projective self consistent
analysis is applied  in the region
where this  method  is valid, i.e. for a bath
with clear   separation of  energy scales.
The formulation in terms of the Landau like functional
\ref{landau} is  useful, because it
allows the introduction of {\it trial} states, which
can be analyzed more easily than the solution of the full
non linear dynamical mean field equations \ref{mf}.

We end the section by recapitulating the logic leading to
the inequality \ref{ineq}. We start with the Mott insulator
and slowly decrease the value of U.
In section \ref{uc1sec} we asked the question when is the
insulating solution unstable to a small perturbation.
The answer is an implicit  equation for U, eq. \ref{uc1}.
Its solution, which we denote $U_{c1}$ is certainly  bigger or equal
than the value of U at which the gap closes, since when the gap
is closing the insulator is definitely unstable against small 
perturbations.
Then we consider a
trial state corresponding to an impurity model in an intermediate
coupling regime.
The condition for instability against this specific
trial state, is given by   eq \ref{uc2}.
While the specific value of  $ < {\vec S} .\sloc >  $
depends on the trial state,
the fact that  the  Mott insulator is unstable  at a value
of U strictly larger than Uc1  and therefore posessing
a finite gap,
relies only on the fact
that $ < {\vec S} .\sloc > <  0$. This is  always true 
for impurity models  which are in the   intermediate coupling 
regime.
The trial state we used gives us {\it a lower bound} to the true value
of $U_{c2}$, since improving the scaling function might result in
a better trial state which can destabilize  an insulator  at an even
larger value of U.   (This optimization
was  done numerically in  ref \cite{moeller1}).

We therefore have proved that an upper bound for the U
which is needed for gap closure is strictly smaller than
a  lower bound  for the U  at which the insulator
becomes unstable against metalization.

Within the context of the Landau analysis we  do  not
have to  worry  about the normalization of $ \delta  \Delta_{L}(\omega) $.
To 
exhibit  the instability against a  normalized trial state,
one can proceed as we did on 
page \pageref{normalize}.

Notice the  two significant differences with respect to the 
previous section:

a) The cost of forming the required Weiss field 
(while still given by  the first term in eq. \ref{projective} and therefore
having an identical form to that considered in the previous section
( see eq. \ref{cost}))
 is now   
 of order $Z$ (and not $Z^2 $ as in the previous section )
 because  the height of the spectral function at the origin is of order unity. 

b) The gain in energy from  the free energy of the impurity model
 is  now also of order $ Z $
This is not surprising because each individual  term  
in the Hamiltonian of the Kondo impurity is  of order Z.

The difference between  the determination of    $U_{c1}$   and $U_{c2}$
can then be traced to the different behavior of the impurity 
model
in the weak coupling and the intermediate coupling regime.
To illustrate this idea    in a 
much simpler setting, we
analyze a very simple  free energy function
in Appendix A.

\section{ Implications and Outlook for  Finite Dimensions }
\label{conclusions}

Our analysis so far was confined to zero temperature.
However the proof of  existence of two coexistent solutions
(one metallic and one insulating)
in a finite interval of interaction strength
at zero temperature, has immediate consequences for  the finite
temperature phase diagram.
Away from the phase boundaries,
the Landau functional, is a  smooth functional of the
hybridization function. 
Since we have two coexistent
solutions, at a finite distance of each
other \cite{distance}, at zero temperature,
by continuity the two solutions persist at finite temperature.

At finite temperatures there is no qualitative difference between
the metal-like  and the insulating-like solution.
Just the low frequency density
of states is quantitatively different (large in the metal, small in the
insulator). 
The coexistence of two solutions
at zero temperature results in a first order phase transition
at finite temperatures 
 since entropy favors the paramagnetic
insulating solution.

In the  temperature interaction strength
phase  diagram of the frustrated Hubbard model
refs \cite{werner} \cite{werner2} \cite{g2}  
\cite{g3} the
points $U_{c1}$ and $U_{c2}$ are  the end points
of  two lines    
$U_{c1}(T) $ and $U_{c2}(T)$   which cross at a finite
temperature second order point at $(U_{MIT}, T_{MIT})$.

Rozenberg et. al. \cite{phase} observed that the $U_{c1}(T) $
and $U_{c2}(T)$ can be continued above $T_{MIT}$, where they
become crossover lines.
The physical interpretation \cite{phase}, as crossover lines, was remarkable
similar, to that of the zero temperature points where they originated.
At  $U_{c1}(T) $ the gap becomes comparable to the temperature
and activation results. At $U_{c2}(T)$ the  Fermi liquid coherence
disappears \cite{phase}.

The limit of  full frustration is rare.
However,
it turns out that for reasons having to do with
orbital degeneracy,
crystal structure, and longer range multi-spin interactions,
many three dimensional transition metal oxides
undergoing a metal to insulator transition, are very nearly frustrated,
and finite temperatures easily stabilize the paramagnetic insulator
phase.

The continuation of the zero temperature
paramagnetic metal to  paramagnetic insulator
transition to finite temperatures,
then becomes immediately relevant for answering a
long standing question of whether
one  can account for the
phase diagram of $Ni Se_{2-x} S_x$ or of $V_2 O_3$ 
in purely  electronic terms without including
explicitly the coupling to the lattice. The answer
suggested in refs
\cite{werner} \cite{g2} and  \cite{phase} was positive, provided one includes
a sufficient degree of magnetic  frustration.

The original  discussion
relied  on approximate methods such as the IPT, QMC,
or exact diagonalization, which were  open to criticism.
We have stressed  in this article
that at zero temperature
there are two coexisting solutions
in a range of interaction strengths  U,
which allows us to conclude  (without recourse to approximations),
that the qualitative features found in references \cite{g2}  \cite{phase}
in IPT
are not artifacts of the approximation, but genuine
features of the solution of the dynamical mean field equations
\ref{mf}.
The character of the phase diagram, is a consequence of the
analytic structure of the free energy functional. One can view,
the different approximation methods, as providing different  approximate
values to the coefficients of the Landau Ginzburg  functional of the order
parameter. As long as the  functional has the correct analytic structure,
the qualitative features of  the solution are preserved (of course, approximate
methods can only give
approximate values of  $T_{MIT}$, $U_{c1}(T) $ $U_{c2}(T)$ etc ). 
Approximations 
such
as the fully self consistent skeleton method
fail  because they do not capture
the correct analytic properties of the free energy functional,
i. e. they  miss  
the paramagnetic
insulating phase.
 
The dynamical mean field solution of the Hubbard model
has  allowed us to study and understand the {\it finite
temperature} consequences of the existence of a Brinkman
Rice quantum critical point
\cite{rice} $U_{c2}$, where  spatial coherence is lost.
This  quantum critical
points is  always
inaccessible because  no system is fully frustrated and
some form of magnetic order always sets in before reaching
the critical point  at zero temperature. The DMFT  has been 
of extraordinary value in revealing the hidden origin
of the anomalous temperature dependence \cite{phase} of many properties.
The {\it observable} anomalous finite
temperature properties, of the 
paramagnetic
phases close to the 
Mott transition point, are connected, via DMFT, to the presence
of a zero temperature 
quantum
critical point.

Having established the physical  {\it relevance} of the paramagnetic metal
to paramagnetic insulator transition in magnetically frustrated systems, we turn to the broader and more interesting question of how reliable
 the DMFT description of this transition is.
 
Conventional mean field theories of classical
phase transitions are qualitatively valid from infinite
dimensions down to a finite upper critical dimension.
The situation is very different in dynamical mean field theory.
As currently formulated, in the paramagnetic insulating phase, 
this theory omits a physically
relevant scale, the magnetic exchange energy.
The dynamical mean field picture of metalization, as a result,
cannot capture the competition between 
intersite magnetic correlations and
the Kondo effect.

In a simplistic picture of the DMFT,
which views  the central site as  a  strongly correlated electron  f, 
and the neighbor as a non interacting bath  electron c,
the f site will  have a well formed moment $S$ ( which captures  the localized
character of the electron) while  the nearby
sites are represented by  a conduction band
which is narrowed to represent the correlated but  
itinerant character of the original  electrons.
Since the bath is uncorrelated, it can only describe a moment,
by becoming infinitely narrow. This does not happen in reality,
as a result of magnetic correlations, which in three dimensions,
ultimately trigger some form of magnetic order.
The qualitative success of the dynamical mean field approach
depends crucially on the weakness of
the  intersite magnetic correlations compared 
to the renormalized Fermi  energy
${\epsilon_F}^*$.
 
Since in the dynamical mean field theory,   the effective Fermi energy
vanishes at zero temperature and at the critical point, the situation
at first sight looks hopeless. 
However  the  experimental 
confirmations of the existence of
an underlying  quantum critical point at $U_{c2}$  
arise from the finite temperature behavior.
As long as the  degree of
 frustration  and the temperature
is sufficiently high so that the relevant magnetic energy
scale is smaller than the temperature, (or when the
distance from the critical point is sufficiently large so  that
the renormalized Fermi energy 
is larger than the magnetic exchange),
we are in a situation where many of the qualitative predictions
of the dynamical
mean field theory are applicable.
some of these predictions
have  actually been  observed experimentally \cite{phase}\cite{shen}.
Underlying these successes
 are some unique features  of the   dynamical mean field theory
which are not shared by ordinary mean field theories.
A single dynamical mean field solution can contain
information about a large number of energy scales. As  the
corresponding  dynamical
mean field theory of spin glasses, it works well in frustrated situations. 
This should be contrasted with static mean field theories, which usually
do not work well when there are several competing ground states.

We now turn to  the outstanding open issue of  how 
the findings of the dynamical mean field theory are 
modified when one is at sufficiently low temperatures,
or sufficiently close to the transition so that the 
magnetic correlations have to be taken into account.
The  considerations  
presented in this note  suggests
significant  departures from the results of  
ynamical mean field theory.

In infinite dimensions, near $U_{c1}$ we have complete
decorrelation of ${\vec  S} $ and
${\vec S_L}
={{\cll}^\dagger} {\vec \sigma} \cll$. Near $U_{c2}$ however
we have a finite value of $ <{\vec  S} . \locspin >$ 
because a band which is infinitely narrow,
can be polarized without any cost, and we succeed in
getting some finite energy from the {\it spin correlations}
between neighboring sites all the way to the transition.

This can not be accomplished
once we have a finite exchange among nearby sites (an effect
which is formally $1/d$ and therefore invisible in the large d paramagnetic
phase but becomes of order one once we allow for magnetic long range order) .
Now the Kondo interaction between the central spin
${\vec S}$
 and
${\vec S_L}$
will 
compete with the exchange interactions with the 
 nearby spins.
As a result the free energy becomes a smoother function
of the hybridization.  
If there is some kind of Kondo effect going on at the
transition it will  now  occur in  a regime of  weaker  coupling strength.
The same conclusion  can also be
reached  from the point of view of the itinerant
electrons. 
If we now accept that the self energy can contain
k dependence the pinning condition $\rho(0) = \frac{1}{D}$ is now
eliminated. The possible Weiss fields can  now be more like
those considered  in the discussion 
in section \ref{uc1sec}.

For these reasons  
we believe that   the destruction of the insulator near 
$U_{c1}$ which 
we described  in section \ref{uc1sec} 
may play a more fundamental role
in finite dimensions or in  partially frustrated situations.

Stated in a different language the large d picture is
one where $\epsilon_F =Z t $,  the renormalized Fermi energy,
 is larger
than $J_{ij}$,  the  magnetic superexchange between a pair of spins.
Clearly  this is justified in the formal infinite d limit because
$J_{ij} =t_{ij}^2/U \propto   {1 \over d} $ is vanishingly small no matter how small
$\epsilon _F$  is.
That situation changes immediately once $J_{ij}$  is finite. 
Clearly new physics is expected when 
$J_{ij}$ is larger than  $ \epsilon _{F}$.

{\it Within DMFT} 
the influence of magnetic correlations on the Mott transition
has been recently studied recently \cite{chitra} 
by approaching the metal insulator transition from an ordered state.
In this case, one can quantitatively study how the magnetic
correlations dramatically modify the behavior near the metal 
insulator transition point. 
In the absence
of magnetic long range order, the task is more difficult.
We have to do it without the guidance of
the infinite d limit , because all those effects
disappear in that case.
These effects appear  next order in $1/d$, but
a complete treatment  of those corrections, is still
missing \cite{progress}.

We  believe  that  
the  basic idea of this paper, i.e.  that   different types
of  Mott transition should be viewed as different
types of bifurcations, of a system of functional equations
for a  metallic order parameter closely
related to the one particle Green's function,  will  also
be useful  beyond the limit of large lattice coordination.

\section{Appendix A}
To illustrate the main features of the
Landau Functional we discuss in this Appendix a toy free
energy which has a structure similar to
eq. \ref{landau}
but which  is much easier to analyze.
This free energy  is  a function of two
	variables :  j
 (which should be thought of as one of the $ {V_k}^2$  of
	the previous sections)  and t which should be thought of as a single
	$\epsilon_k$. It  is given by: 
\begin{equation}
F[t,j]=
  {\frac{{j^2}}{t}} + (t - 
  {\sqrt{{j^2}\,{{{\Js}}^2} + {t^2}}}
)
-{\frac{10\,j\,{\Js}}{{e^{{\frac{t}{j}}}}}} 
\label{toy}
\end{equation}
The analogy with the discussion in
	previous sections should be apparent, the first term is the cost
	of making the Weiss field, the second term
	is the energy that the impurity gains once
	the Weiss field is formed . We have
	mimicked the non analytic dependence of the free energy on the coupling 
strength in weak coupling by the exponential term which does not have a Taylor
 series in j, around j=0
The analysis leading to the determination of $U_{c1}$ consists of 
simply evaluating 
whether there is
	an energy gain by increasing j away from zero, in weak coupling, that is for 
a fixed value of t and for an infinitesimal value of j.
	This stability analysis is equivalent to checking whether the second 
derivative of the free energy with respect to j, at fixed t is positive or 
negative at the origin
{}From equation \ref{toy} we  conclude 
that the $U_{c1}$ like instability occurs at the value of U ($\Js$ is a
 function of U ) such that 
	 
\begin{equation}
{\frac{2}{t}} - {\frac{{{{\Js}}^2}}
    {{\sqrt{{t^2}}}}}=0
\end{equation}
We now perform the determination of the critical value which parallels
Uc2, i.e. we perform a minimization of the free energy without assuming
whether the minimum occurs at weak or strong coupling. In other words,
we focus on the functional which tells us if energy is gained or lost by
increasing w , the weight at the origin, for fixed value of the 
inverse coupling
strength a which we define as  $a = {t \over j}$.
$U_{c2}$ is determined by then optimizing with respect to
the coupling strength  a
.
	 
\begin{equation}
f[j= w, t= w. a]= g[w,a]=
{\frac{w}{a}} + a\,w - {\frac{10\,{\Js}\,w}
    {{e^a}}} - {\sqrt{{a^2}\,{w^2} + 
      {{{\Js}}^2}\,{w^2}}}
\end{equation}
 Now we notice that since in this toy model  the free
energy  is  linear in w, the instability occurs when
this term is negative (w by definition is always positive).
The calculation that parallels $U_{c2}$ requires us 
to minimize with respect to a and search 
for instability. Terms non linear
in w could be added to the toy free energy
so as to  obtain well defined solutions to the full non linear 
problem, but   will  not be considered   here.
 
It is  clear that at  $ U_{c2}$, the system prefers to be metallic and 
favors a 
strong coupling state.
\section{Appendix B}
To derive eq. (\ref{uc2})   we need the asymptotic behavior of the
real part of the Green's function in 
the Anderson impurity model in the regime $ZD   \ll  \omega  \ll  \Delta_g$, 
(i.e. for energies much less than the Mott Hubbard gap but greater
than the resonance width ). 
From the asymptotic behavior
\begin{equation}
{\delta \Delta_L} (i \omegan) \approx \frac {D^2 Z} {4 i \omegan}
\end{equation}
it follows that $\int_\infty^\infty \phi_1 (x) dx= {1 \over 4}$
\def\wn{{\omega_n}}
The low energy part of the f electron operator is given in equation 
\ref{fop}.
\def\irl{ \int \rho_{L}( \epsilon ) d \epsilon}
\def\sums{\sum_{\sigma}}
Using the definition  of the local electron  spin
\ref{locspin}
and  the identity
\def\vecs{\vec S }
\def\vsigma{\vec  \sigma}
\begin{equation}
{1 \over 2} \sums < \{ (\vecs
{c^\dagger}_{L} \vsigma  {)_\sigma}   ,  ({\vec S } \vsigma c_{L} {)_\sigma} \}>
=2  (\frac{3}{8} - < \vecs . \sloc  >)
\label{b2}
\end{equation}
we obtain
\def\nombre{{Z \over 2}  {(D J_{s})}^{2}
(\frac{3}{8} - < \vecs . \sloc  >)}
\def\nombret{{1 \over 2}  {(D J_{s})}^{2}
(\frac{3}{8} - < \vecs . \sloc  >)}
\begin{equation}
\irl= \frac{1}{2} \sum_{\sigma} <\{F_{\sigma} {{F^\dagger}_{\sigma}}\} >
\approx {\nombre }
\end{equation}
It follows that $\int_\infty^\infty \phi (x) dx= {\nombret}$
which is then used in   eq. \ref{uc2}  in the text.

NOTE ADDED:
After this paper was  completed,
two  numerical studies 
of the problem discussed in this note
have appeared. In an algorithmic tour de force, Bulla  \cite{bulla}
obtained a value of   
$U_{c2}   $  in excellent agreement with  earlier results 
\cite{moeller1}.
Using a different numerical method
Gebhard and Noack \cite{dresden},
obtained  a value of  $U_{c1}   $
which is 
in  excellent agreement with  earlier work  \cite{phase}
\cite{g3}. 
Their value for $U_{c2}$ however, is   much smaller than 
all previous estimates.

ACKNOWLEDGMENTS:
My understanding of this problem was shaped by fruitful
collaborations with  
R. Chitra, V. Dobrosavlevic,  D. Fisher,  A Georges, H. Kajueter, W. Krauth,
E. Lange,  G. Moeller, G. Palsson
M. Rozenberg, 
Q. Si and X.Y. Zhang.
E. Abrahams, A. Georges, A. Millis, M. Rozenberg, A. Ruckenstein,   Q. Si and D. Vollhardt
made useful comments
on several early versions of this manuscript.
I am particularly   grateful to E. Lange for   numerous
discussions   and for
a careful proofreading of the manuscript. 
Finally this article would not have been written without Philippe
Nozi\`eres insistence that the energetics of the Mott transition
deserved clarification.
This work was supported by NSF DMR 95-29138.


\begin{thebibliography}{99}
\bibitem{caveat} The linear terms in eqs \ref{cost} and
\ref{gain} can be shown to vanish as the
support of the perturbation $\dl$ gets concentrated
in the low energy region.
\bibitem{caveat3}
The hybridization function is  really a non local order parameter. 
It measures,
the stiffness or sensitivity to boundary conditions  of the single particle
Green's
function, which in large dimensions 
is simply related to the low energy conductivity.
\bibitem{caveat2}
It was shown in  refs \cite{moeller2}  \cite{g3}
that in fact the gap closes at $U_{c1}$.
\end{thebibliography}
\end{document}